\documentclass[a4paper,fleqn,usenatbib]{mnras}
\usepackage{ulem}

\usepackage{xcolor}
\usepackage{multirow}
\usepackage{graphicx}
\usepackage{booktabs}
\usepackage{threeparttable}
\usepackage[T1]{fontenc}

\DeclareRobustCommand{\VAN}[3]{#2}
\let\VANthebibliography\thebibliography
\def\thebibliography{\DeclareRobustCommand{\VAN}[3]{##3}\VANthebibliography}

\usepackage{graphicx}	
\usepackage{amsmath}	
\usepackage{amssymb}	

\title[Searching for chemical anomalies in HD\,196067-68]{Elemental abundances differences in the massive planet-hosting wide binary HD\,196067-68\thanks{Based on observations made with ESO Telescopes at the La Silla Paranal Observatory under programme ID 0101.C-0275, 0101.C-0275, 0103.C-0785, 188.C-0265, and 088.C-0323.}}

\author[Flores et al.]{
M. Flores$^{1,2,3}$\thanks{E-mail: matiasflorestrivigno@conicet.gov.ar},
J. Yana Galarza$^{8}$,
P. Miquelarena$^{1,2,3}$,
C. Saffe$^{1,2,3}$,
M. Jaque Arancibia$^{6,7}$,
\newauthor
R. V. Ibañez Bustos$^{4}$,
E. Jofré$^{3,5}$, 
J. Alacoria$^{1,3}$, and 
F. Gunella$^{1,3}$
\\
$^{1}$Instituto de Ciencias Astron\'omicas, de la Tierra y del Espacio (ICATE), Espa\~na Sur 1512, CC 49, 5400 San Juan, Argentina.\\
$^{2}$Facultad de Ciencias Exactas, F\'isicas y Naturales, Universidad Nacional de San Juan, San Juan, Argentina.\\
$^{3}$Consejo Nacional de Investigaciones Cient\'ificas y T\'ecnicas (CONICET), Argentina.\\
$^{4}$Laboratoire Lagrange, Université Côte d’Azur, Observatoire de la Côte d’Azur, CNRS, Boulevard de l’Observatoire, CS 34229, 06304 Nice Cedex 4, France\\
$^{5}$Universidad Nacional de Córdoba -- Observatorio Astronómico de Córdoba, Laprida 854, X5000BGR, Córdoba, Argentina\\
$^{6}$Instituto de Investigación Multidisciplinar en Ciencia y Tecnología, Universidad de La Serena, Raúl Bitrán 1305,
La Serena, Chile.\\
$^{7}$Departamento de F\'isica y Astronom\'ia, Universidad de La Serena, Av. Cisternas 1200, La Serena, Chile\\
$^{8}$The Observatories of the Carnegie Institution for Science, 813 Santa Barbara Street, Pasadena, CA 91101, USA.
}

\date{Accepted XXX. Received YYY; in original form ZZZ}

\pubyear{2015}

\begin{document}
\label{firstpage}
\pagerange{\pageref{firstpage}--\pageref{lastpage}}
\maketitle

\begin{abstract}
It has been suggested that small chemical anomalies observed in planet-hosting wide binary systems could be due to planet
signatures, where the role of the planetary mass is still unknown. We search for a possible planet signature by analyzing
the $T_{C}$ trends in the remarkable binary system HD\,196067--HD\,196068. At the moment, only HD\,196067 is known to host a planet which is near the brown dwarf regime. 
We take advantage of the strong physical similarity between both stars, which is crucial to achieving the highest possible precision in stellar parameters and elemental chemical
abundances. This system gives us a unique opportunity to explore if a possible depletion of refractories in a binary
system could be inhibited by the presence of a massive planet. We performed a line-by-line chemical differential study, employing
the non-solar-scaled opacities, in order to reach the highest precision in the calculations. After differentially comparing both stars, HD\,196067 displays a 
clear deficiency in refractory elements in the $T_{C}$ plane, a lower iron content (0.051 dex) and also a lower Li {\sc i} content (0.14 dex) 
than its companion. In addition, the differential abundances reveal a $T_{C}$ trend. These targets represent the first cases
of an abundance difference around a binary system hosting a super-Jupiter. Although we explored several scenarios to explain
the chemical anomalies, none of them can be entirely ruled out. Additional monitoring of the system as well as studies of larger sample of wide binary systems hosting massive planets, are needed to
better understand the chemical abundance trend observed in HD\,196067-68.
\end{abstract}




\section{Introduction}
Small chemical anomalies could be originated from the planet formation process, as suggested by \citet[][hereafter M09]{2009ApJ...704L..66M}. Implementing a differential abundance analysis in a sample of 11 solar twins, M09 found that the Sun is deficient in refractory elements when compared to the average abundance of this group. In addition, a clear correlation between the condensation temperatures ($T_{C}$) of the elements and the differential abundances was observed. The authors argued that the refractories\footnote{In this work, we adopted as refractory and volatile species those with $T_{C}$ > 900 $\mathrm{K}$ and $T_{C}$ < 900  $\mathrm{K}$, respectively.}, missing in the solar atmosphere, are locked in the rocky planets, asteroids and rocky cores of giant planets. Although these results have been validated by other authors \citep[e.g.,][]{2010A&A...521A..33R, 2011ApJ...737L..32S,2014A&A...561A...7R,Bedell:2018ApJ...865...68B,2020MNRAS.495.3961L}, other studies suggest that the $T_{C}$ trends, also observed in some binary systems, could also be a result of the galactic chemical evolution (GCE), dust cleansing, dust-gas segregation, and planet engulfment events \citep[e.g.,][]{2011A&A...528A..85O,2014A&A...564L..15A,2015A&A...579A..20M,2015A&A...579A..52N,2015A&A...582L...6S,2015ApJ...804...40G,2017A&A...604L...4S,2018A&A...620A..53G,2021NatAs...5.1163S,2023A&A...676A..87H}. Recently, \citet[][]{2020MNRAS.493.5079B}, using evolutionary models for protoplanetary disks, associated the sequestration of refractory elements to the formation of Jupiter-like planets, showing that 
the origin of the small abundance
depletions is yet subject of a prolific debate.

Binary systems are the ideal astrophysical laboratories to test several processes. For instance, they can be key objects for testing the stellar evolution theory \citep[e.g.,][]{1999ApJ...521..297I,2019MNRAS.482.1231J}. These objects are also important to carry out some stellar activity studies such as the stellar planet interactions (SPI's), and also the search for prolonged activity minimum states similar to the solar Maunder Minimum (MM) \citep[][]{2014A&A...565L...1P,2021A&A...645L...6F}. Moreover, one of the most important applications is to test the star-planet connection since the possible chemical signature, imprinted as a consequence of several processes such as the planet formation, planet engulfment event, and gas-dust segregation, can be better tested by a strictly differential study in those binary systems where one of the components hosts an exoplanet. The assumed co-eval and co-natal (i.e., same age and chemical composition) nature allows us to neglect the GCE, and even other environmental effects \citep[see][for details]{2016ApJ...819...19T} which could mimic a $T_{C}$ trend. The atomic diffusion can be ruled out only when both stars of the binary system are twins. As shown in \citet[][]{2021MNRAS.508.1227L} through a detailed study of chemical abundances of binary stars, the atomic diffusion effects are non-negligible when the components of the system present large differences in $T_{\rm{eff}}$ and $log\,g$ ($\Delta T_{\rm{eff}} > 200\, \mathrm{K}$ and $\Delta log$ g $>$ 0.07 dex, respectively). In addition, the more similar the components are to each other, the higher precision in the differential abundance determination is achieved.

Unfortunately, up to now reported studies in binary systems with similar components (where at least one of them harbours a planet) are still scarce and their resulting abundances determinations show a different behaviour with the $T_{C}$. For example, the binary systems HAT-P-1, HD\,80606-7, and HD\,106515 \citep{2014ApJ...785...94L,2015A&A...582A..17S,2019A&A...625A..39S} do not show a clear $T_{C}$ trend, in contrast with the 16 Cyg, WASP-94, HAT-P-4, and WASP-160 systems, where a $T_{C}$ trend was detected when both components were differentially compared \citep[see][for more details]{2014ApJ...790L..25T,2016ApJ...819...19T,2017A&A...604L...4S,2019A&A...628A.126M,2021AJ....162..291J}. In addition, some differences have also been found in the chemical composition of binary systems where both stars host planets, for example XO-2 \citep[see][for more details]{2015ApJ...808...13R,2015A&A...583A.135B} and HD\,133131AB  \citep[][]{2016AJ....152..167T}. The results mentioned above have given rise to different scenarios about their origin showing that further high-precision studies of planet-hosting binary systems are needed to shed light over these clues.

In order to investigate the possible chemical signature of the planet formation through a strictly differential analysis, we have focused on the study of wide binary systems where at least one of the components has a planet or a debris-disk \citep[e.g.,][]{2019A&A...625A..39S,2021AJ....162..291J}. Moreover, as a result of these studies, we have also been able to improve the precision in the calculation of stellar parameters and chemical abundances by using non-solar-scaled opacities, being this method particularly useful for the analysis of the $T_{C}$ trends \citep[see][for more details]{2018A&A...620A..54S}.

The binary system composed by HD\,196067 ($=$HIP 102125, $V=6.44$, $B-V=0.61$, hereafter component A), and HD\,196068 ($=$HIP 102128, $V= 7.21$, $B-V=0.64$, hereafter component B) is a remarkable target located at 44 $\pm$ 5 pc from the Sun with a separation of 16.8$^{''}$ \citep[corresponding to $\sim$932 au,][]{2004ApJS..150..455G}. Their reported spectral types are G0V\,$+$\,G1V, respectively \citep{2006AJ....132..161G}. The A component harbours a massive companion (12.5$_{-1.8}^{+2.5}M_{jup}$) close to the planet-brown dwarf boundary \citep[][]{2021AJ....162..266L}, detected by the CORALIE planet-search survey employing RV technique. It has an orbital period of 3638 days and is located at 5.02 au \citep[][]{2013A&A...551A..90M}. In contrast, there is no planet detected orbiting around the B component.

A close inspection of wide binary systems hosting planets \citep[][]{2011ApJ...740...76R,2014ApJ...785...94L,2015A&A...582A..17S,2017A&A...604L...4S,2019A&A...625A..39S,2021AJ....162..291J} reveals a number of remarkable features that justify the election of HD\,196067-68 for the present study. On one side, the possible role of the planetary mass is unknown. Mutual chemical differences have been found for binary systems hosting planets with a mass lower than 3 $M_{jup}$ \citep[e.g.][]{2014ApJ...790L..25T,2017A&A...604L...4S,2021AJ....162..291J}. However, no chemical differences were found between the stars of the system HD\,80606-7 \citep[][]{2015A&A...582A..17S,2016ApJ...818...54M} nor HD\,106515 \citep[][]{2019A&A...625A..39S}, both hosting massive planets (4.0 $M_{jup}$ and 9.0 $M_{jup}$, respectively). 
The possibility that chemical differences disappear for massive planets, suggests that the planetary mass could play a role in the differences observed. This would require studying additional systems, particularly those having massive planets. The detection of a massive companion (12.5 $M_{jup}$) hosted by the binary system HD\,196067-68, converts this system into an excellent target for this aim, being the most massive planet detected to date. In addition, the study of the $T_{C}$ trends is a challenging task that requires the highest possible precision in the derivation of stellar parameters and abundances. In this case, the strong physical similarity between both components of HD\,196067-68 ($\Delta T_{\rm{eff}}$ $\sim$39 $\mathrm{K}$ and $\Delta log$ g $\sim$0.06 dex) is an important advantage for this twin pair through the use of a line-by-line differential analysis. Finally, the period and mass of the planet hosted by HD\,196067 makes it one of the very few systems where the scenario of \citet[][]{2020MNRAS.493.5079B} could be tested.

At the moment, both components of this system have been only taken into account in some statistical studies of chemical abundances \citep[e.g.][]{2018A&A...614A..55A,2018A&A...612A..93M,2021NatAs...5.1163S}. In particular, \citet{2021NatAs...5.1163S} studied a sample of 107 binary systems to test the planet engulfment scenario. To do so, the authors carried out a high-precision abundance determination. As a result, they detected chemical signatures of planetary ingestion in $\sim$25\% of their sample, reporting the binary system HD\,196067-68 as chemically anomalous ($\Delta$Li {\sc i} \,$\sim$0.16 dex and $\Delta$[Fe$/$H] $\sim$0.09 dex). Nevertheless, there is no in-deep exploration of this remarkable binary system, including a more exhaustive analysis of the chemical anomalies and possible $T_{C}$ trends. Therefore, a high-precision determination of stellar parameters and abundances in this pair of twin-stars could help to clear up the origin of the possible $T_{C}$ trends in the high-planetary mass regime, which is poorly studied to date.

This work is organised as follows: In \S 2, the observations and data reduction are described. In \S 3, our high-precision chemical abundances analysis is presented. 
In \S 4, we show our results and discussion. Finally, our main conclusions are provided in \S 5.

\section{Observations and data reduction}

The stellar spectra of HD\,196067 and HD\,196068 were obtained from the European Southern Observatory (ESO) archive\footnote{\url{http://archive.eso.org/wdb/wdb/adp/phase3_spectral/form?phase3_collection=HARPS}}. These data were acquired with the HARPS spectrograph\footnote{\url{http://www.eso.org/sci/facilities/lasilla/instruments/harps/overview.html}} installed at the 3.6 m La Silla telescope (ESO) in Chile \citep{2003Msngr.114...20M}. The resolving power of this fibre-fed instrument is $R$ $\sim$115\,000 and the spectra cover a range from 3782 \AA \, to 6913 \AA.

The observations were taken between 2018 and 2019 with the same  
instrumental configuration, where the second component 
was observed immediately after the first one (HD\,196067). The resulting  
signal-to-noise (S/N) for each component is $\sim$400, measured near 6600 \AA.
In addition a solar spectrum with similar S/N, which is necessary for our differential study, was obtained by observing the Vesta asteroid. All these spectroscopic data were reduced with the HARPS pipeline\footnote{\url{http://www.eso.org/sci/facilities/lasilla/
instruments/harps/doc.html}}, which carries out the usual reduction tasks including
bias subtraction, flat fielding, sky subtraction, order extraction, and
wavelength calibration. These spectra were combined, normalized, and radial velocity corrected by using the software package IRAF\footnote{IRAF is distributed by the National Optical Astronomical Observatories, which is operated by the Association of Universities for Research in Astronomy, Inc., under a cooperative agreement with the National Science Foundation.}.

\section{Fundamental atmospheric parameters and abundances analysis}

As in previous works \citep[e.g.,][]{2019A&A...625A..39S,2021AJ....162..291J}, the stellar atmospheric parameters ($T_{\rm{eff}}$, $log\,g$, [Fe$/$H], and $v_{turb}$) of HD\,196067 and HD\,196068 were first determined by applying the full line-by-line differential technique\footnote{By ``full'' we mean that line-by-line differences were considered in both the derivation of stellar parameters and (not only) abundances.}, using the Sun as the reference star and adopting the typical solar atmospheric parameters ($T_{\rm{eff}}$= 5777 $\mathrm{K}$, $log\,g$= 4.44, [Fe$/$H]=0, and $v_{turb}$= 1.0 $\mathrm{km\,s}^{-1}$). Solar abundance values were taken from \citet{2009ARA&A..47..481A}. However, a final solar $v_{turb}$ of 1.1 $\mathrm{km\,s}^{-1}$ was derived after requiring a zero slope in the absolute abundances of Fe {\sc i} lines as a function of the reduced equivalent width ($EW_{r}$). It is important to mention that the parameters of the stars A and B were calculated through the non-solar-scaled method. As shown in \citet{2018A&A...620A..54S}, the use of non-solar-scaled opacities gives as a result smaller differences in both stellar parameters and chemical abundances in comparison to the classical solar-scaled approach.

Before searching for the ionization and excitation balance for the Fe {\sc i} and Fe {\sc ii} lines, we first manually measured the equivalent widths ($EWs$) of these and the other chemical elements employing the splot task of IRAF. The chemical elements list, together with several laboratory data (wavelength, excitation potential, oscillator strengths, and damping constant), were obtained from  
\citet{2014MNRAS.442L..51L}, \citet{2014ApJ...791...14M}, and \citet{2014ApJ...795...23B}. Then, the program FUNDPAR \citep{2015A&A...582A..17S,2018A&A...620A..54S} was used to force the ionization and excitation balances. For this purpose, this differential version, which internally uses the MOOG\footnote{We employed the current MOOG release (November 2019, \url{https://www.as.utexas.edu/~chris/moog.html}).}  program \cite[][]{1973ApJ...184..839S} together with atmospheres models of ATLAS12  \citep{1993KurCD..13.....K}, requires minimizing the slopes of the differences in Fe {\sc i} and Fe {\sc ii} abundances vs. the excitation potential and the $EW_{r}$.

\begin{figure}
\centering
\includegraphics[scale=0.425]{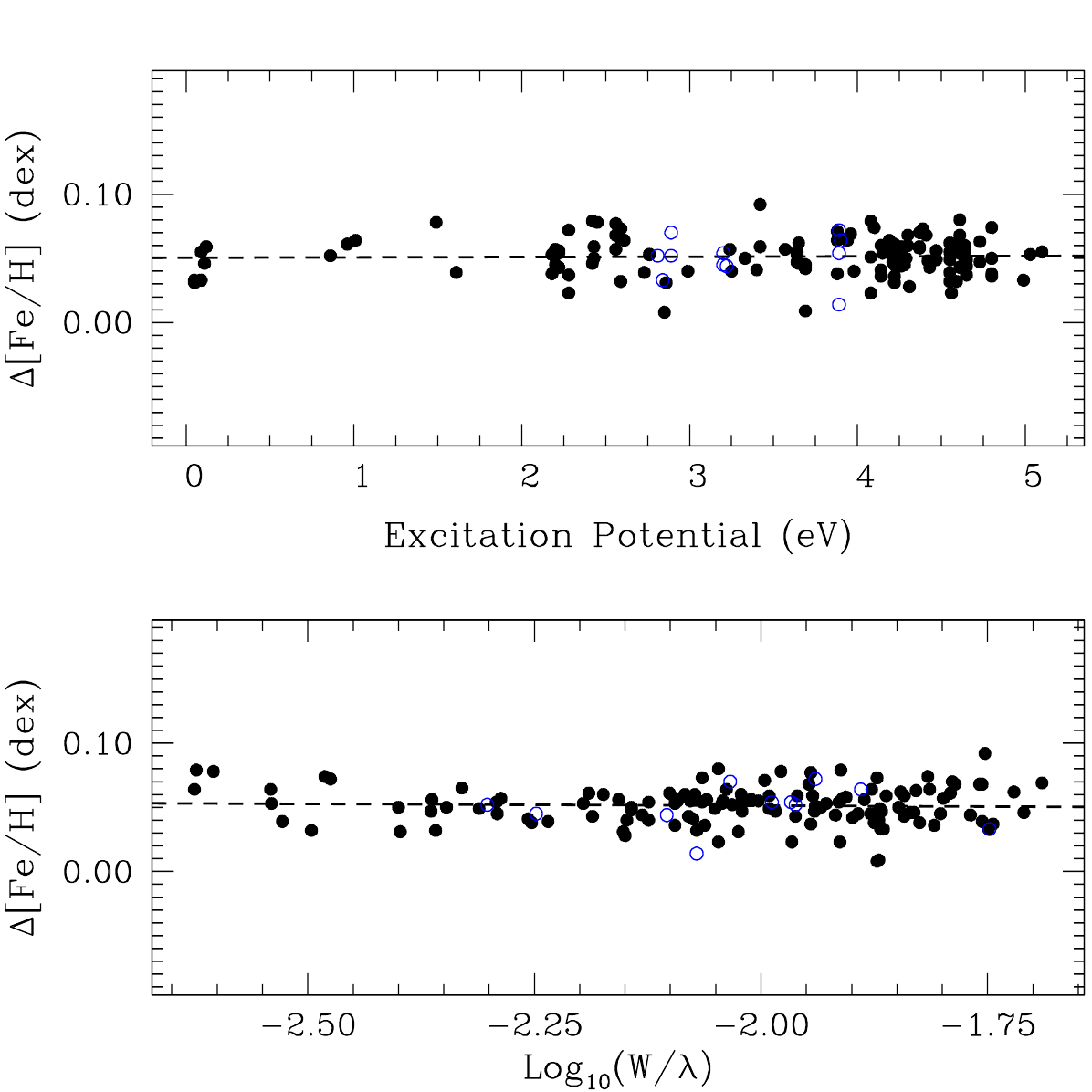}
\caption{Differential iron abundance as a function of the excitation potential (upper panel) and the reduced equivalent width (lower panel) of the B component relative to the A (i.e., B$-$A). Fe {\sc i} and Fe {\sc ii} lines are represented with filled and blue empty circles, respectively.  A linear fit to the data is also shown in dashed lines.}     
\label{Figthree}
\end{figure}

We summarize our results in Table \ref{wt}. For the target HD\,196067 (A$-$Sun, in this case), the stellar parameters and its uncertainties are:
$T_{\rm{eff}} = 6059 \pm\,52$ $\mathrm{K}$, $log\,g= 4.26 \pm\,0.12$ dex, [Fe$/$H]= 0.216 $\pm\,0.009$ dex, and $v_{turb}=1.48 \pm\,0.07$ $\mathrm{km\,s}^{-1}$. While for HD\,196068 (B$-$Sun), the following parameters were obtained: $T_{\rm{eff}} = 6020 \pm\,52$ $\mathrm{K}$, $log\,g= 4.32 \pm\,0.10$ dex, [Fe$/$H]= 0.267 $\pm\,0.009$ dex, and $v_{turb}=1.41 \pm\,0.07$ $\mathrm{km\,s}^{-1}$. The errors in stellar parameters have been calculated by considering both the individual and the mutual covariance terms of the error propagation \citep[see][for more details]{2015A&A...582A..17S}. Then, the stellar parameters for the B star were recalculated using the A component as reference (i.e., B-A). For this aim, the parameters adopted for the A component were taken from the solution (A-Sun). In contrast with the previous step, where the Sun was used as reference (i.e., B$-$Sun), now it can be noted that the uncertainties have been reduced: $T_{\rm{eff}}$ = 6020 $\pm\,42 \, \mathrm{K}$, $log\,g= 4.32 \pm\,0.08$ dex, [Fe$/$H]= 0.267 $\pm\,0.005$ dex, and $v_{turb}=1.41 \pm\,0.06$ $\mathrm{km\,s}^{-1}$. In Figure \ref{Figthree}, we compare the Fe {\sc i} and Fe {\sc ii} abundances to the excitation potential and the $EW_{r}$ of the B star, using its companion (A component) as reference. It can be clearly seen that
the B component is more metal-rich than its companion by $\sim$0.05 dex.

In order to explore the nature of this system, we estimated the age ($\tau$), mass (M$_{\star}$) and radius (R$_{\star}$) for both components using the Yonsei-Yale (Y$^{2}$) isochrones of stellar evolution \citep{Yi:2001ApJS..136..417Y, Demarque:2004ApJS..155..667D} through the q$^{2}$ package \citep{Ramirez:2014A&A...572A..48R}, which uses the spectroscopic $log\,g$ (4.087 and 4.352, respectively) and absolute magnitudes from \textit{Gaia} DR3 parallaxes to increase the precision \citep[see][]{Spina:2018MNRAS.474.2580S,Yana-Galarza:2021MNRAS.504.1873Y}. Consequently, we obtained $\tau = 3.50^{+0.65}_{-0.27}$ Gyr, $M_{\star} = 1.330^{+0.013}_{-0.013}$\ M$_{\odot}$, and $R_{\star}$ = $1.730 ^{+0.044}_{-0.039}$\ R$_{\odot}$ for the A component, and $\tau = 2.10^{+1.03}_{-0.34}$ Gyr, $M_{\star} = 1.190^{+0.017}_{-0.018}$\ M$_{\odot}$, and $R_{\star}$ $= 1.190 ^{+0.034}_{-0.010}$\ R$_{\odot}$ for the B component (see Table \ref{wt} for details).

\begin{table}
\begin{scriptsize}
\caption{Stellar parameters derived in this work.}
\begin{tabular}{|p{1.5cm}|p{1.5cm}|p{1.5cm}|p{1.5cm}|} 
\toprule
\toprule
\multicolumn{1}{|c|}{(Star-reference)} & \multicolumn{1}{c|}{(A-Sun)} & \multicolumn{1}{c|}{(B-Sun)} & \multicolumn{1}{c|}{(B-A)} \\
\hline
T$_{eff}$ ($\mathrm{K}$)   & 6059 $\pm$	52  & 6020 $\pm$52 & 6020 $\pm$42 \\  
$log\,g$  (dex) & 4.26 $\pm$	0.12 & 4.32 $\pm$ 0.10  & 4.32 $\pm$ 0.08  \\
$\mathrm{[Fe/H]}$  (dex) & 0.216 $\pm$ 0.009  & 0.267 $\pm$ 0.009  & 0.267 $\pm$ 0.005  \\
$v_{turb}$ ($\mathrm{km\,s}^{-1}$) & 1.48  $\pm$ 0.07  & 1.41 $\pm$ 0.07  & 1.41 $\pm$ 0.06  \\
$\tau$   (Gyr) & $3.50^{+0.65}_{-0.27}$  & $2.10^{+1.03}_{-0.34}$  & \,\, \, \,\ \,\, \,  --- \\
M$_{\star}$ (M$_{\odot})$ & $1.330^{+0.013}_{-0.013}$  & $1.190^{+0.017}_{-0.018}$ & \,\, \, \,\ \,\, \,  --- \\
R$_{\star}$ (R$_{\odot})$ & $1.730^{+0.044}_{-0.039}$  & $1.190^{+0.034}_{-0.010}$  & \,\, \, \,\ \,\, \,  --- \\
\bottomrule
\end{tabular}
\label{wt}
\end{scriptsize}
\end{table}

After the calculation of the fundamental parameters, we derived the abundances for 32 chemical species by using the non-solar-scaled method. For the case of V  {\sc i}, Mn {\sc i}, Co {\sc i}, Cu {\sc i}, Y {\sc ii},  Ba {\sc ii}, Eu {\sc ii}, and Li {\sc i} elements, the hyperfine structure splitting (HFS) was considered. Consequently, these abundances were measured through spectral synthesis with the program SYNTHE \citep{1981SAOSR.391.....K} by using the HFS constants of \citet{1995KurCD..23.....K}. In Table \ref{tab} we present the final differential abundances of the B component when its companion (A) is considered as the reference star. In addition, and for comparative purposes, we include the differential abundances of the A and B components relative to the Sun. We also included an estimation of the total error $\sigma_{tot}$ for each chemical specie. In this way, the total uncertainty was derived by quadratically adding the observational error $\sigma_{obs}$ (obtained as $\sigma /  \sqrt{(n-1}))$\footnote{Being $\sigma$ the standard deviation of the different lines and n the number of lines.}, the internal error associated with uncertainties in the stellar parameters $\sigma_{par}$\footnote{These values have been obtained after quadratically adding the abundance variation when modifying the stellar parameters by their uncertainties.}, and the error from [Fe/H]. For those chemical species with only one spectral line, i.e., Gd {\sc ii}, Sr {\sc i}, Sr {\sc ii}, and Li {\sc i}, we adopted the average standard deviation of the other elements as the observational error. This same criteria has been applied in other works \citep[e.g.,][]{2019A&A...625A..39S,2020A&A...641A.145S}. Non-NLTE corrections were considered for Na {\sc i}, Mg {\sc i}, and Li {\sc i} species by using the INSPECT\footnote{Data obtained from the INSPECT database, version 1.0 (\url{www.inspect-stars.com}} tool \citep{2012MNRAS.427...27B,2012MNRAS.427...50L}.

\begin{table*}
\caption{Line-by-line chemical abundances for A and B components relative to the Sun (A$-$Sun, B$-$Sun), and B relative to A (B$-$A).}
\centering
\begin{threeparttable}
\begin{tabular}{|l|r|r|r|r|r|r|r|r|r|r|r|r|}
 \hline
 \hline
\multicolumn{1}{c}{} & \multicolumn{4}{c}{A$-$Sun} & \multicolumn{4}{c}{B$-$Sun} & \multicolumn{4}{c}{B$-$A} \\
\cmidrule(rl){2-5} \cmidrule(rl){6-9} \cmidrule(rl){10-13}
{Species} & {[X/Fe]} & {$\sigma_{obs}$} & {$\sigma_{par}$} & {$\sigma_{tot}$} & {[X/Fe]} & {$\sigma_{obs}$} & {$\sigma_{par}$} & {$\sigma_{tot}$} & {[X/Fe]} & {$\sigma_{obs}$} & {$\sigma_{par}$} & {$\sigma_{tot}$}\\
\midrule
C  {\sc i} &  $-$0.005      & 0.031 & 0.045 &       0.056                      &  $-$0.053              & 0.020 & 0.041 &    0.048                         & $-$0.064 & 0.011 & 0.023   &0.026 \\
Na {\sc i} & 0.041          & 0.007 & 0.018 &       0.023                      &  0.034                 & 0.005 & 0.018 &    0.023                         & $-$0.021 & 0.003 & 0.011   &0.014 \\
Mg {\sc i} &  0.084         & 0.017 & 0.029 &       0.036                      &  0.112                 & 0.008 & 0.030 &    0.033                         & 0.015 & 0.011 & 0.018      &0.023 \\
Al {\sc i} &  0.030         & 0.003 & 0.024 &       0.028                      &  0.087                 & 0.008 & 0.025 &    0.029                         & 0.037 & 0.006 & 0.016      &0.018 \\
Si {\sc i} &  0.023         & 0.004 & 0.004 &       0.014                      &  0.031                 & 0.004 & 0.004 &    0.013                         & $-$0.001 & 0.003 & 0.002   &0.008 \\
S  {\sc i} &  $-$0.007      & 0.008 & 0.040 &       0.043                      &  $-$0.040              & 0.026 & 0.038 &    0.048                         & $-$0.047 & 0.018 & 0.021   &0.028 \\
Ca {\sc i} &  0.003         & 0.004 & 0.014 &       0.019                      &  $-$0.004              & 0.005 & 0.014 &    0.019                         & $-$0.005 & 0.006 & 0.008   &0.012 \\
Sc {\sc i} &  0.036         & 0.013 & 0.029 &       0.034                      &   0.045                & 0.018 & 0.029 &    0.036                         & 0.000 & 0.019 & 0.019      &0.027 \\
Sc {\sc ii} &  0.054        & 0.007 & 0.023 &       0.027                      &   0.081                & 0.009 & 0.019 &    0.025                         & 0.018 & 0.009 & 0.009      &0.015 \\
Ti {\sc i} &    0.013       & 0.004 & 0.010 &       0.017                      &   0.037                & 0.004 & 0.010 &    0.017                         & 0.019 & 0.003 & 0.006      &0.010 \\
Ti {\sc ii} & 0.008         & 0.005 & 0.016 &       0.021                      &   0.016                & 0.005 & 0.014 &    0.019                         & 0.003 & 0.004 & 0.007      &0.010 \\
V  {\sc i} & 0.018          & 0.015 & 0.032 &       0.038                      &   0.036                & 0.012 & 0.035 &    0.040                         & 0.016 & 0.004 & 0.021      &0.023 \\
Cr {\sc i} & $-$0.015       & 0.004 & 0.009 &       0.016                      &   $-$0.011             & 0.005 & 0.009 &    0.016                         & 0.006 & 0.004 & 0.005      &0.009 \\
Cr {\sc ii} & $-$0.011      & 0.010 & 0.027 &       0.032                      &   $-$0.019             & 0.014 & 0.024 &    0.030                         & $-$0.006 & 0.008 & 0.012   &0.017 \\
Mn {\sc i} &  $-$0.013      & 0.028 & 0.045 &       0.055                      &   $-$0.021             & 0.042 & 0.050 &    0.067                         & $-$0.011 & 0.014 & 0.032   &0.035 \\
Co {\sc i} &  0.051         & 0.007 & 0.015 &       0.021                      &   0.072                & 0.005 & 0.015 &    0.020                         & 0.010 & 0.002 & 0.009      &0.012 \\
Ni {\sc i} & 0.041          & 0.002 & 0.005 &       0.013                      &   0.056                & 0.002 & 0.005 &    0.013                         & 0.005 & 0.002 & 0.003      &0.008 \\
Cu {\sc i}  &  0.040         & 0.037 & 0.037 &       0.053                      &   0.040                & 0.014 & 0.036 &    0.041                         & $-$0.021 & 0.014 & 0.022   &0.027 \\
Zn {\sc i} & 0.008          & 0.033 & 0.025 &       0.043                      &   0.027                & 0.013 & 0.023 &    0.029                         & $-$0.030 & 0.002 & 0.017   &0.018 \\
Sr {\sc i} &  $-$0.042      & 0.018 & 0.052 &       0.056                      &   $-$0.054             & 0.019 & 0.054 &    0.058                         & 0.023 & 0.013 & 0.033      &0.036 \\
Sr {\sc ii} & 0.005         & 0.018 & 0.046 &       0.050                      &   $-$0.038             & 0.019 & 0.040 &    0.045                         & $-$0.008 & 0.013 & 0.020   &0.024 \\
Y  {\sc ii}& $-$0.087      & 0.007 & 0.064 &       0.066                      &   $-$0.122             & 0.007 & 0.074 &    0.076                         & $-$0.001 & 0.014 & 0.064   &0.065 \\
Ba {\sc ii} & $-$0.196      & 0.007 & 0.070 &       0.072                      &   $-$0.246             & 0.023 & 0.074 &    0.077                         & $-$0.006 & 0.007 & 0.052   &0.053 \\
La {\sc ii} & $-$0.126      & 0.032 & 0.051 &       0.062                      &   $-$0.136             & 0.055 & 0.043 &    0.072                         & 0.023 & 0.023 & 0.022      &0.032 \\
Ce {\sc ii} & $-$0.082      & 0.011 & 0.026 &       0.031                      &   $-$0.111             & 0.016 & 0.022 &    0.030                         & 0.002 & 0.006 & 0.012      &0.015 \\
Sm {\sc ii} & 0.019         & 0.057 & 0.053 &       0.079                      &   0.005                & 0.012 & 0.044 &    0.048                         & $-$0.003 & 0.069 & 0.023   &0.073 \\
Eu {\sc ii} &  $-$0.022     & 0.141 & 0.041 &       0.147                      &   $-$0.036             & 0.163 & 0.042 &    0.168                         & $-$0.006 & 0.021 & 0.044   &0.049 \\
Gd {\sc ii} & $-$0.021      & 0.018 & 0.051 &       0.055                      &   $-$0.040             & 0.019 & 0.044 &    0.049                         & $-$0.011 & 0.013 & 0.022   &0.026 \\
Dy {\sc ii} & $-$0.142      & 0.006 & 0.053 &       0.055                      &   $-$0.126             & 0.018 & 0.045 &    0.050                         & 0.026 & 0.012 & 0.024      &0.027 \\
\hline
Li\textsuperscript{*} {\sc i} & 2.39 & 0.018 & 0.051 & 0.055 & 2.53 & 0.019 & 0.052 & 0.056 & 0.140 & 0.013 & 0.042 & 0.044\\
 \hline
 \hline
\end{tabular}
\label{tab}
\begin{tablenotes}
\item[] Notes:
\item[] For final differential abundances (columns 2, 6, and 10),  with the exception of lithium, we have adopted the standard notation, ie., [X/Fe] = [X/H] - [Fe/H].
\item[*] Absolute abundances of lithium (A(Li)).
\end{tablenotes}
\end{threeparttable}
\end{table*}

\section{Results and discussions}
\begin{figure}
\centering
\includegraphics[width=\hsize]{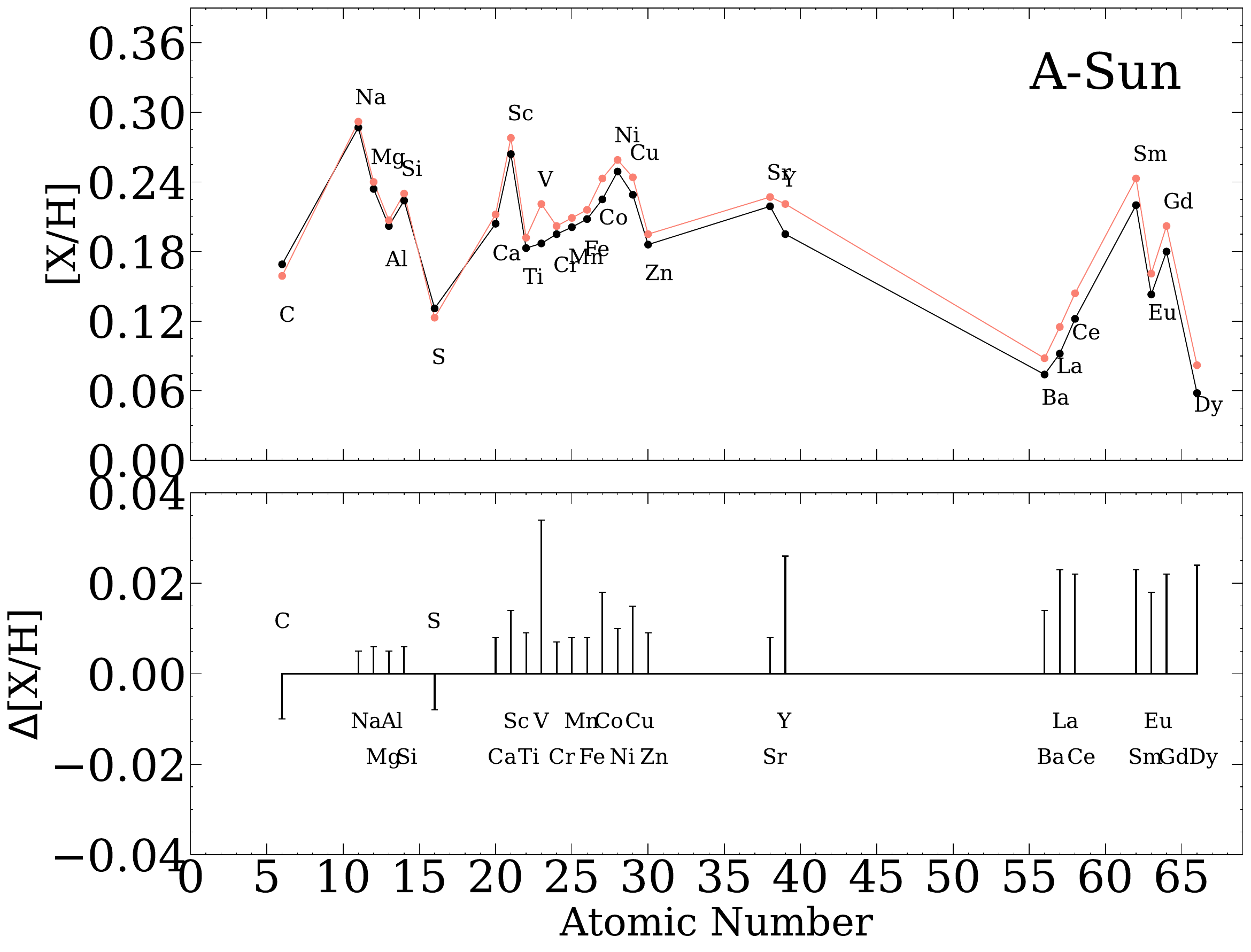}
\caption{Solar-scaled (black points line connected) vs. non-solar-scaled (red points line connected) method. Differential abundances of (A$-$Sun) as a function of atomic numbers are presented in the upper panel. The differences between both methods are shown with bars in the lower panel.}
\label{Figfour}
\end{figure}

\begin{figure}
\centering
\includegraphics[width=\hsize]{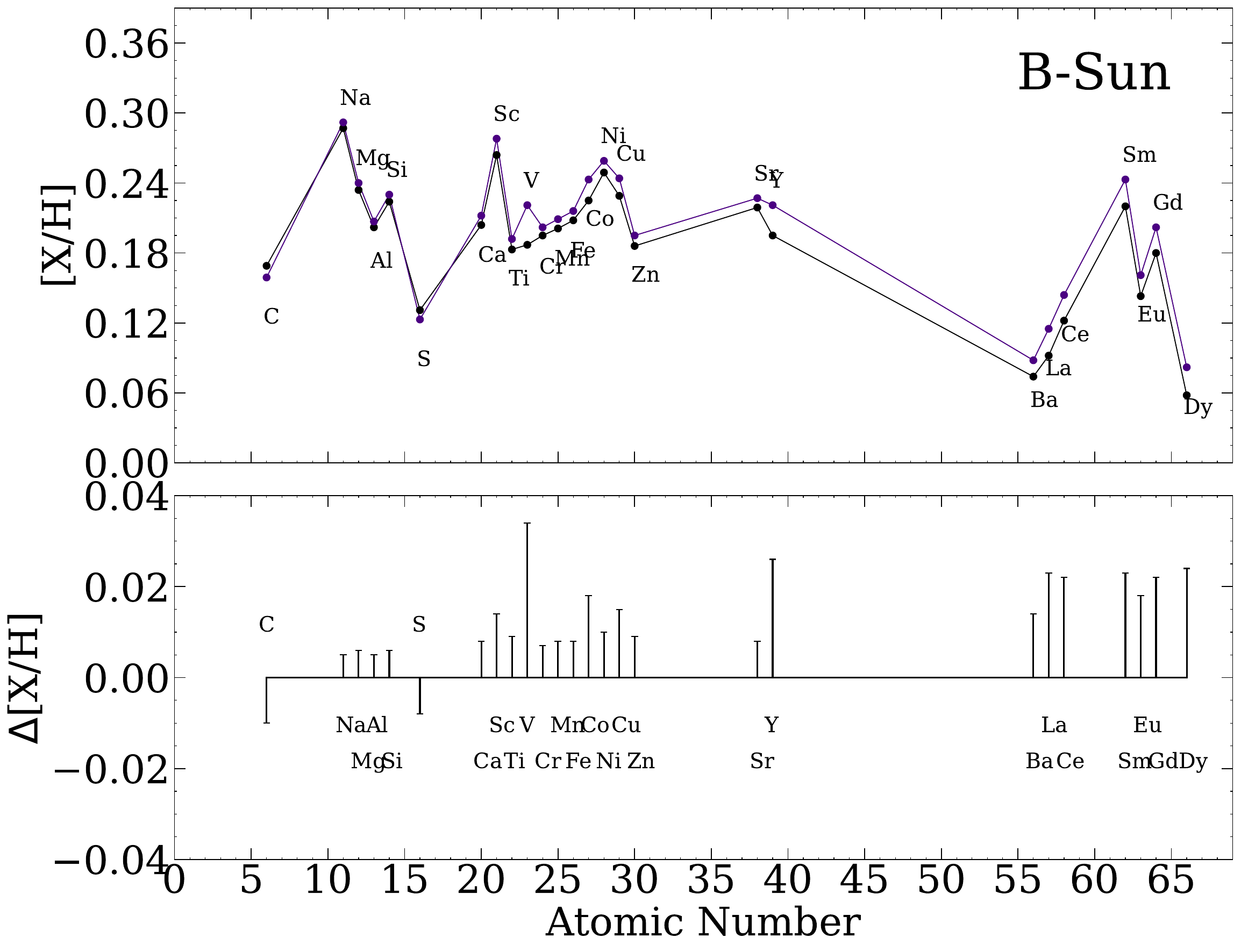}
\caption{Similar to Figure \ref{Figfour}, but for B$-$Sun.}
\label{Figfive}
\end{figure}

The improvement of the  precision in a line-by-line strictly differential study could be crucial for the detection of a possible chemical signature of planet formation. In this line, \citet{2018A&A...620A..54S} showed that the application of the non-solar-scaled method is an appropriate choice when a mutual comparison of different chemical species is needed. Figures \ref{Figfour} and \ref{Figfive} show the differential abundances obtained with the non-solar-scaled method (red and purple points, respectively) as a function of the atomic numbers for the A and B components using the Sun as reference. For comparison, the stellar abundances derived from the classical method have been also included (black points). Differences of the order of $\sim$0.03 dex can be observed for some elements in the lower panels of both figures (for example V, Y, Sm, Eu, Gd, and Dy). As can be noted, the differences between both methods are of the order of NLTE corrections or GCE effects \citep[see][for more details]{2018A&A...620A..54S}.

These abundances were corrected by GCE effects using the [X/Fe]-age correlation of \citet{Bedell:2018ApJ...865...68B} and following the prescription given by \cite{Spina:2016A&A...585A.152S} and \cite{Yana_Galarza:2016A&A...589A..65G}. As the age of the pair is lower than the solar age, we added GCE trends to its chemical composition so that this is also corrected to the same age as the Sun (see columns two and six reported in Table \ref{tab}.).

\begin{figure}
\centering
\includegraphics[width=\hsize]{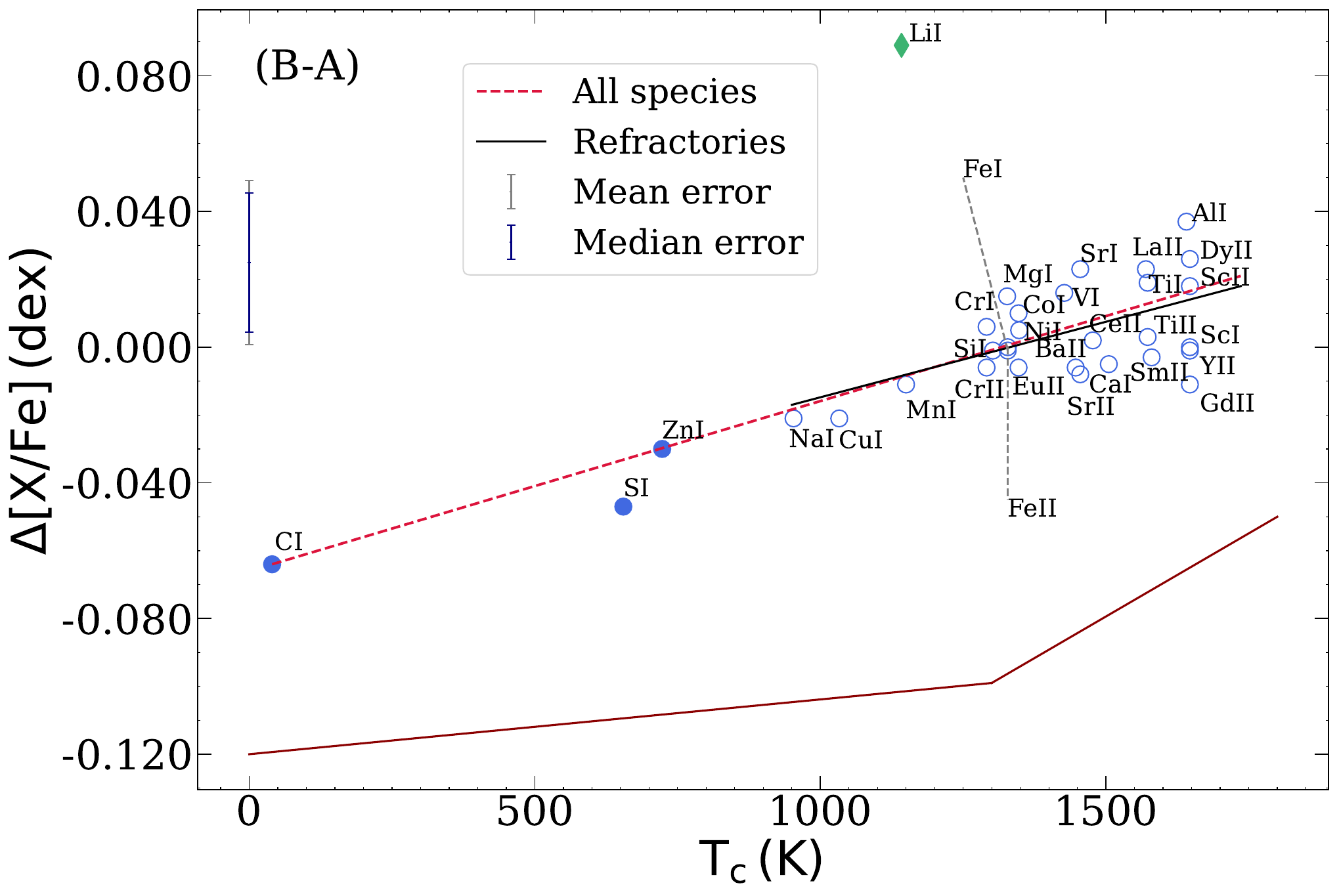}
\caption{Differential abundances between B and A (B$-$A) as a function of the $T_{C}$ for volatiles (filled blue circles) and refractory elements (unfilled blue circles). The linear fit to all the elements is indicated with the red dashed line, while the fit to refractory elements alone is indicated with the black continuous line. The statistical mean and median of the abundances are shown in the upper left corner. The solar-twin trend of M09 corresponds to the continuous fit shown in red. The differential Li {\sc i} abundance is shown with a green diamond. The position of Fe {\sc i} and Fe {\sc ii} is also indicated.} 
\label{Figeight}
\end{figure}

Better precision in the determination of the stellar abundances could be achieved if we compare both stars to each other. In particular, their similar initial chemical composition due to a probable common natal environment \citep{2007AJ....134.2272R,2012MNRAS.421.2025K,2012Natur.492..221R} and the similarity in stellar parameters of this twin-binary system, allow us to diminish the GCE effects and reduce the dispersions in our calculations. In Fig. \ref{Figeight} we plotted the differential abundances as a function of the $T_{C}$ for the B component relative to A. For comparative purposes, we included the solar-twins trend of M09 (in red), the linear fit to all the elements (red dashed line) and the refractories species (dark continuous line). The green diamond corresponds to the Li {\sc i}. For this particular element, the differential abundances of line 6707.80 \,\AA \, were determined by using spectral synthesis. Its position in this plot deserves particular attention (see the next section), mainly because the  
content of Li {\sc i} has been linked to the accretion scenario \citep[e.g.][]{1999MNRAS.308.1133S,2002ApJ...572.1012S,2018ApJ...854..138O,Soares_Furtado_2021}. Note that the error bars of each element ($\sigma_{tot}$) are not shown in the figure, however, they are reported in Table \ref{tab}.

Then, for the purpose of quantifying a possible $T_{C}$ trend in Fig. \ref{Figeight}, we calculated the corresponding slopes\footnote{This calculation includes the uncertainties in the abundance determinations ($\sigma_{tot}$) presented in the column 13 of the Table \ref{tab}.}, the correlation coefficient $\rho$, and their corresponding dispersions ($\sigma_{s}$ and $\sigma_{\rho}$, respectively) after applying  a weighted linear fit. In addition, we carried out a statistical test to evaluate the presence of a possible correlation.

 \begin{table}
 \caption{Slopes and dispersions of each fit. The posterior probability distribution of the correlation coefficient with its dispersion, and the corresponding 95$\%$ credible intervals (95$\%$ CI) are also included.}
 \begin{scriptsize}
 \begin{tabular}{|p{2cm}|p{2.2cm}|p{1.3cm}|p{1.3cm}|}
 \hline
 \hline
 B star-A star & slope $\pm$ $\sigma_{s}$ \,[$10^{-5}$ dex/K] &   $\rho \pm \sigma_{\rho}$ &  95$\%$ CI \\
 \hline
 (B-A)$_{all}$   & 4.99 $\pm$ 1.14  & 0.798$\pm$ 0.066 &  (0.668, 0.912)    \\
 (B-A)$_{refrac}$   &  4.62	$\pm$ 1.72 & 0.501$\pm$ 0.131 &   (0.224, 0.734)  \\
  \hline
 \hline
 \end{tabular}
 \label{yt}
 \end{scriptsize}
 \end{table}

The results of the previous analysis are summarized in Table \ref{yt}. Due to the fact that the differential abundance of lithium is outside of the general trend, we have decided to exclude it from the fits. The strength of the correlations was estimated using the Bayesian framework implemented in Python code by \citet{2016OLEB...46..385F}. As can be viewed, this analysis indicates a significant slope for the general and the refractory trends (within $\sim$4$\sigma$), including a good correlation coefficient for both cases (0.798$\pm$ 0.066 and 0.501$\pm$ 0.131, respectively). It is important to remark that both slopes (general and refractory) are indistinguishable from each other. These values (4.99 $\pm$ 1.14 $\times \, 10^{-5}$ dex/K and 4.62 $\pm$ 1.72 $\times \, 10^{-5}$ dex/K) suggest a clear $T_{C}$ trend between the stars A and B of this stellar binary system.

At the moment, a number of binary systems have shown chemical anomalies when both stars are mutually compared \citep[e.g.,][]{2016ApJ...819...19T,2017A&A...604L...4S,2018ApJ...854..138O,2019A&A...628A.126M,2020ApJ...888L...9N,2021AJ....162..291J,2021NatAs...5.1163S}. As previously noted, the corresponding planetary masses involved in some of these systems, whose chemical differences also show a correlation
with the $T_{C}$, do not exceed 3 $M_{jup}$. In contrast, the binary systems HD\,80606-7 and HD\,106515A-B, which do host massive planets (4.0 $M_{jup}$ and 9.0 $M_{jup}$, respectively), did not reveal the presence of chemical anomalies \citep[see][for details]{2015A&A...582A..17S,2019A&A...625A..39S}. Then, to
determine if this apparent trend, i.e., if the mutual chemical differences are inhibited in those systems that harbour massive planets, we decided to study the pair HD\,196067-68 which hosts 
a planetary companion of 12.5 $M_{jup}$. Remarkably, as can be observed in Fig. \ref{Figeight}, 
our analysis shows the presence of chemical anomalies when both stars are differentially compared. 
In this way, our results would be indicating that a high planetary mass does not necessarily 
hamper the chemical differences between both components of the binary system.

\subsection{Possible scenarios to explain the $T_{C}$ trend of the twin-binary system HD\,196067-68:}

\subsubsection{Planet engulfment}

The higher metallicity ($\Delta$[Fe$/$H] $\sim$0.05 dex) detected in the B component accompanied by a positive $T_{C}$ trend and also a Li excess ($\sim$0.14 dex), could be signatures of a planet ingestion \citep[e.g.][]{2001ApJ...556L..59P,2017A&A...604L...4S,10.1093/mnras/stz3169,2021AJ....162..291J,2021ApJ...922..129G}. This would be in agreement with the results of \citet[][]{2021NatAs...5.1163S}, who performed a chemical statistical study of 107 binary systems to test this scenario. These authors found evidence in favor of this scenario in $\sim$25\% of their sample, where the binary system HD\,196067-68 is included. In addition, they determined which component of this binary system is chemically anomalous by comparing the refractory content of the stars with that expected from a control sample with similar metallicities. The authors proposed that the engulfment scenario is responsible for both the increase in metallicity and the Li {\sc i} content of the chemically anomalous star. Then, following the scenario of \citet[][]{2021NatAs...5.1163S}, the chemically anomalous star is the component B, having a higher metallicity and Li {\sc i} abundance than the component A (see their Figure 3). Additionally, an estimation of the volatile-to-refractory ratio using our values of [C/Fe] for both components (-0.005 $\pm$ 0.056 and -0.053 $\pm$ 0.048 for the stars A and B, respectively) indicates a higher content of refractory elements in the B component. Then, these data  suggest  that the B component possibly ingested rocky material, in agreement with the scenario of \citet[][]{2021NatAs...5.1163S}.

In order to test if the planet engulfment scenario could have taken place in the system HD\,196067-68, we used the \textsc{terra}\footnote{\url{https://github.com/ramstojh/terra}} code \citep{Yana_Galarza:2016A&A...589A..65G}, which allows us to create synthetic abundances to match with the measured ones. In brief, the code simulates the convective mass of the iron-rich star (B), given its mass and metallicity, to then estimates the amount of rocky material by adding a combination of meteoritic and terrestrial abundance to the convective mass to reproduce the observed abundance pattern. A more detailed explanation about \textsc{terra} can be found in \cite{Yana_Galarza:2016A&A...589A..65G}. The results indicate that the ingestion of $6.2 \pm 1.6 \ M_{\oplus}$ (5.4 of chondrite-like composition plus 0.8 of Earth-like composition) is necessary to reproduce the observed abundance in the B star. The errors of the predicted abundance are computed from the reduced chi-square value. Specifically, it multiplies the reduced chi-square by the appropriate critical values corresponding to a given degree of freedom and the 95\% confidence level to generate a confidence interval representing the uncertainty in abundance estimation. The confidence interval bounds, upper and lower, are not always symmetric. The current convective mass of B is estimated in 0.012 $M_{\odot}$, and it was calculated from a double interpolation between the mass and [Fe/H] of the B component using convective masses from the Y$^{2}$ database \citep{Yi:2001ApJS..136..417Y, Demarque:2004ApJS..155..667D}.

\begin{figure}
\centering
\includegraphics[width=\hsize]{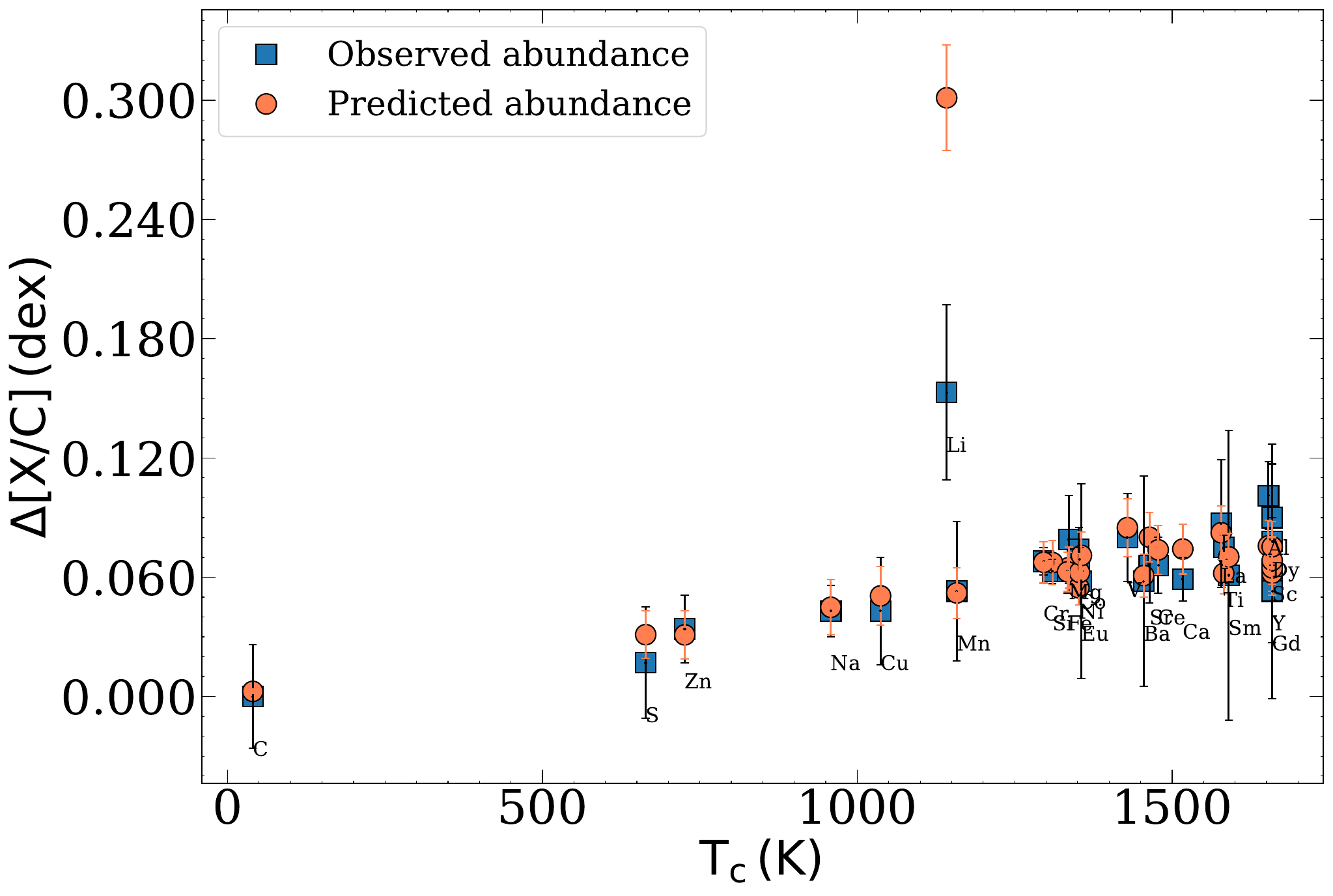}
\caption{Differential chemical abundance of (B-A) as a function of dust condensation temperature ($T_{C}$). The predicted abundances (orange circles) are estimated from an ingestion of  $6.2 \pm 1.6\ M_{\oplus}$. The errors of the predicted abundance are estimated from the reduced chi-square value.}      
\label{Figenine}
\end{figure}

Figure \ref{Figenine} shows a notable agreement between the model (orange circles) and the observed abundances (B$-$A, blue squares) for most elements, except for lithium, which is overpredicted by $\sim$0.14 dex. However, special attention must be paid here, as \textsc{terra} assumes that planetary ingestion has occurred recently and each component had the same initial amount of lithium at the time of accretion. Therefore, the difference in Li between components can be interpreted as a planet engulfment. The overpredicted abundance estimated by the model could indicate that the engulfment occurred before the time of accretion assumed by \textsc{terra}. Hence, our results do not exclude the planet engulfment scenario. Li serves as a reliable indicator of planet engulfment events when they occur roughly at the current age of the iron-rich star, as demonstrated in the case of the binary system HIP\,71726-71737 \citep{Yana-Galarza:2021MNRAS.504.1873Y} where \textsc{terra} precisely predicts the abundance pattern of HIP 71726, including Li. Nonetheless, we are aware that simulating engulfment presents challenges, as it involves considering several variables, including: i) the amount of stellar Li depleted by each component; ii) the timing of the accretion episode; and iii) the convective mass at the time of accretion.

To go a little further than what is generally done in literature, when adopting an engulfment event to explain a $T_{C}$ trend, here we have tried to describe in more detail how and when this event could have occurred in the chemically anomalous B component. To do so, taking into account that no planets have been detected yet orbiting the star B, two strong assumptions need to be invoked. First, considering the strong physical similarity between stars A and B, their high-metallicity content (0.216 dex for A and 0.267 dex for B), and the giant planet detected around star A, it seems plausible to suppose that star B could have formed a giant gas planet as well and possibly rocky material. The second assumption states that the supposed giant planet caused star B to ingest rocky material rich in refractory elements as a consequence of the Kozai effect\footnote{The Kozai mechanism originates oscillations in the eccentricity and inclination of a planet as a consequence of the presence of a remote stellar companion \citep[see][for details]{1962AJ.....67..591K}.}. This mechanism could induce planet migration, potentially resulting in the collision of planets with stars \citep[e.g.,][]{2003ApJ...589..605W,2007ApJ...669.1298F,2012ApJ...754L..36N,2015ApJ...808...14M,2015ApJ...799...27P,2020MNRAS.491.2391C}. In fact, the high eccentricity of several planets in binary systems has been explained through this effect, \citep[e.g.,][]{2003ApJ...589..605W,2021AJ....162..266L}, including  the case of HD\,196067b, whose eccentricity suggests that this mechanism may have played an important role in the dynamics of our binary system HD\,196067-68, affecting the stability of planetary orbits over time. Interestingly, \citet[][]{2003ApJ...589..605W} demonstrated that the migration timescales for HD\,80606b, is at least on the order of $\sim$0.7 Gyr. Therefore, the similarity in mass and separation between HD\,80606-7 ($\sim$$1.1 \ M_{\odot}$ for both components and $\sim$1000 au) and our pair allows us to suppose that the component B (HD\,196068) could have experienced a similar migration timescale. This supports the hypothesis that the engulfment occurred recently, as the \textsc{terra} code predicted. Given the age of star B (2.1 Gyr) and the migration timescale, the engulfment may have occurred at the age of $\sim$1 Gyr, so signatures of engulfment are still detectable. Recently, \citet{2023MNRAS.521.2969B} demonstrated that possible signatures of accretion are largest and longest-lived for 1.2 M$_{\odot}$ stars, observable above 0.05 dex levels for $\sim$5 Gyr after engulfment (see their Fig. 4). Interestingly, the mass of the B component is $\sim$1.2 M$_{\odot}$, with a corresponding enrichment of refractory elements at $\sim$0.055 dex. Therefore, we propose that the migration of a hypothetical giant planet triggered the accretion of refractory material, either from the inner regions of the planetary system or from the giant planet's core itself, resulting in the observed refractory excess in the star HD\,196068. Similar migration scenarios have been proposed for other binary systems \citep[e.g.,][]{2014A&A...572A..49N,Teske_2015,2017A&A...604L...4S,2018ApJ...854..138O,2021AJ....162..291J}. We stress, however, that our hypothesis does not necessarily imply that HD\,196067 has not also experienced an accretion process. If it has, it may have been less effective.

\begin{figure}
\centering
\includegraphics[width=\hsize]{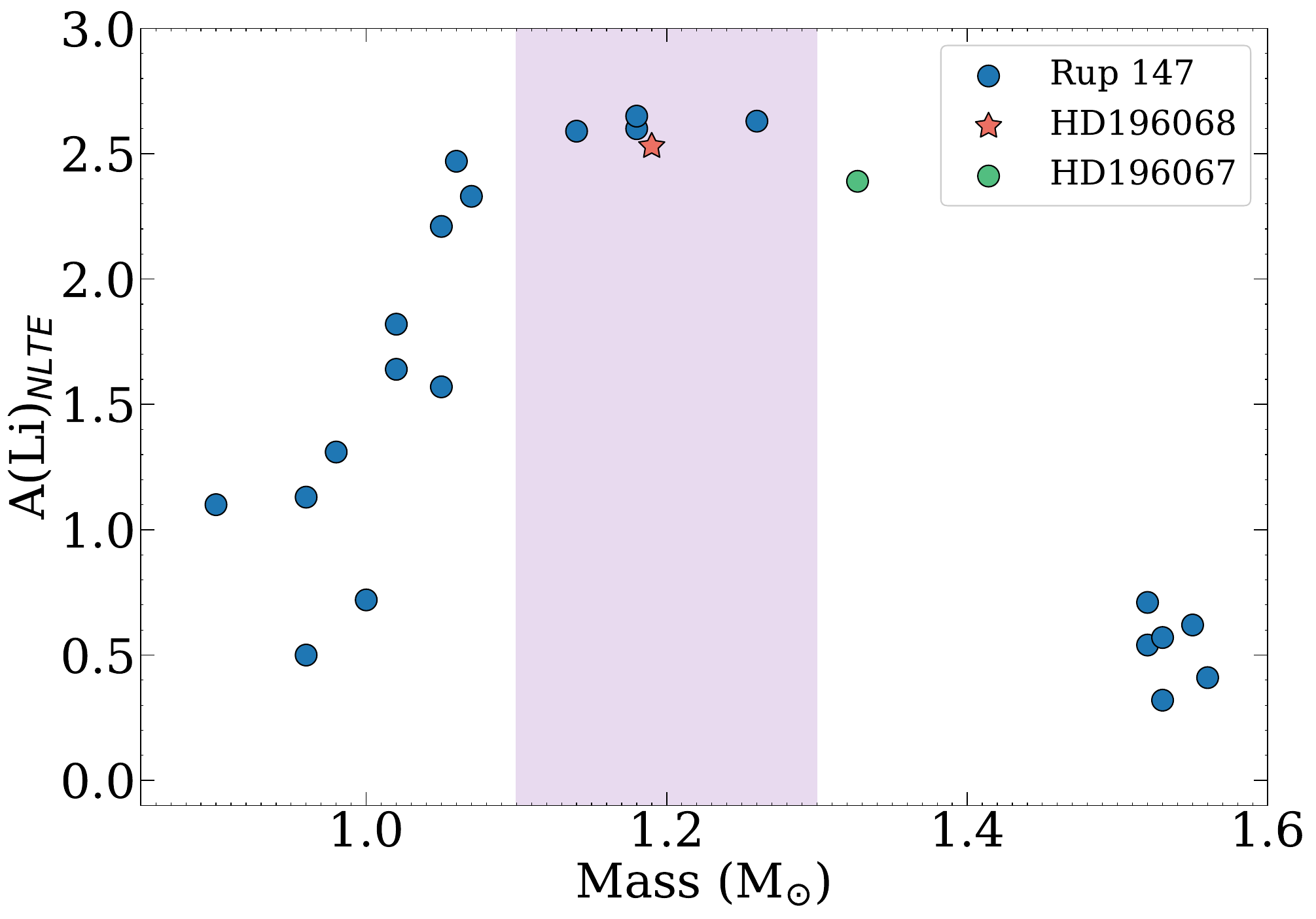}
\caption{A(Li)$_{\rm{NLTE}}$ evolution as a function of stellar mass for the open cluster Ruprecht 147 and the binary system HD\,196067-68. The shadow region represents the lithium plateau \citep{Castro:2016A&A...590A..94C}. The Ruprecht 147 A(Li) data were taken from \citet{Bragaglia:2018AA...619A.176B}.}
\label{Figten}
\end{figure}

Despite the agreement of our
analysis and the predictions obtained with \textsc{terra}, it is noteworthy to mention that the lithium difference observed between both components could also be attributed to differences in their corresponding ages, temperatures, masses and metallicities \citep[e.g.,][]{2005A&A...442..615S,Ramirez:2012ApJ...756...46R, Castro:2016A&A...590A..94C, Bragaglia:2018AA...619A.176B,2023MNRAS.525.4642R}. For instance, taking the open cluster Ruprecht 147 (Figure 5 in \citealp{Bragaglia:2018AA...619A.176B}) as a reference for the lithium evolution of our system due to its similarity in age ($2.5-3.0$ Gyr) and metallicity (0.1-0.23 dex), we can clearly see in Figure \ref{Figten} that the difference in mass between components (0.14 $M_{\odot}$) explains very well the difference in Li (0.14 dex). And the fact that the iron-rich star has more lithium abundance than its companion is possibly because the latter is leaving the lithium plateau (between 1.10 and 1.3 $M_{\odot}$), where this element is barely destroyed during the MS stage independently of stellar mass and metallicity \citep[see][]{Castro:2016A&A...590A..94C}. Beyond the lithium plateau ($>$1.3 $M_{\odot}$), the lithium decreases due to a gravitational diffusion process below the convective zone. Despite the fact that
these results seem to weaken the accretion scenario, the evolution of the Li {\sc i} abundance observed in solar-type stars remains difficult to interpret \citep[see][for more details]{2017A&A...597A..34M,Baraffe_2017,Soares_Furtado_2021}.

\subsubsection{Planet locking}
As a result of unprecedented precision work performed in the Sun and 11 solar twins, M09 concluded that the depletion of refractory elements shown by the Sun together with a strong correlation with the $T_{C}$ could be attributable to the planet formation process. In this line, the primordial 
content of refractory elements in the Sun was possibly affected by the creation of both the terrestrial planets and/or the cores of giant planets, at the time of star and planet formation. This depletion of refractory elements in the Sun's photosphere has been also confirmed and related to signatures of the planet formation by other authors \citep[e.g.,][]{2010A&A...521A..33R,2014A&A...561A...7R,2018A&A...618A.132K,Bedell:2018ApJ...865...68B}. In particular, by means of a detailed chemical study in a larger sample (79 Sun-like stars), \citet{Bedell:2018ApJ...865...68B} also found the deficiency in refractory material of the Sun. In addition, the M09 scenario 
has also been invoked to explain the chemical anomalies observed in a number of binary systems \citep[e.g.,][]{2016A&A...588A..81S,2016A&A...589A..17Y,2020MNRAS.495.3961L,2021AJ....162..291J}.

Then, the positive $T_{C}$ trend in refractory elements observed in Fig. \ref{Figeight} could be explained under the scenario of M09. When comparing mutually both stars (i.e., B$-$A), the B component is more metal-rich than its companion ($\sim$0.051 dex, as shown in Fig. \ref{Figthree}). In addition, the average values of the differential abundances [X/H] are higher for refractory elements ($\sim$0.055 dex) when compared to volatile ones ($\sim$0.004 dex). In other words, the planet locking in this binary system could be related 
to a rocky planet (not yet detected) or with the core of the super-Jupiter planet (12.5$_{-1.8}^{+2.5}M_{jup}$, \citealp[][]{2021AJ....162..266L}) around the A component, similar to the case of the binary system 16-Cyg \citep[][]{2014ApJ...790L..25T}.

\subsubsection{Gas-dust segregation}

It has been suggested that a $T_{C}$ trend can also be explained under the gas-dust segregation scenario \citep[][]{Gaidos_2015}.
Based on this idea, recently \citet[][]{2020MNRAS.493.5079B} performed an evolutionary model to test if, at the moment of a giant planet formation, the gas-dust segregation process operating at a protoplanetary disk could produce a chemical fingerprint. In this work, the authors demonstrated that a forming giant planet (i.e., Jupiter analogues) can open gaps \footnote{The presence of these gaps has been revealed by ALMA and VLT/SPHERE observations \citep[e.g.,][]{2018ApJ...863...44A}} which would allow the gas to be accreted into the proto-star. In contrast, a substantial mass of exterior dust would be sequestered outside of the planets orbits without the possibility of being accreted. This could lead to a refractory element deficiency on the order of 0.02-0.06 dex, similar to the chemical pattern shown by the Sun (0.04 dex). 

The results of \citet[][]{2020MNRAS.493.5079B} were recently reproduced by \citet[][]{2023A&A...676A..87H}. In this work, the authors performed a numerical simulation to study how the evolution of the accretion disks could affect the stellar abundances, including a better description for the individual elements 
that participate in the accretion process as well as 
a more comprehensive treatment of planetary growth by pebble accretion.
In addition, they compared the observational abundances of the wide binary system 
HD\,106515 with those derived from their model, concluding
that the elemental abundance difference showed by this system can be originated as a consequence of the formation of both a giant planet and also planetesimals.

Then, taking into account the above mentioned scenario, the deficit of refractory elements showed by the A component ($\sim$0.05 dex relative to its companion B) was possibly produced through a gas-dust segregation mechanism at the early formation of its distant giant planet HD\,196067b (5.02 au, 12.5$_{-1.8}^{+2.5}M_{jup}$). However, at the moment, these models do not predict a theoretical $T_{C}$ trend to properly compare with the observations.

\section{Conclusions}

We carried out a detailed line-by-line differential analysis of the twin binary system HD\,196067-68. Fundamental parameters and stellar abundances of both components were obtained employing both the classical and the non-solar-scaled methods, being the latter a better approach accomplished for the first time in this binary system. These calculations were first performed by using the Sun as reference star (i.e., A$-$Sun and B$-$Sun), and then the A component as reference (i.e., B$-$A). We used this last  
best solution to search for possible $T_{C}$ trends comparing the binary system to the solar-twins trend of M09. As a result, when both stars are compared to each other (B$-$A), a clear different chemical composition is revealed where the differential abundances show a higher content of refractories compared to volatiles. In addition, all chemical elements, as well as the refractory species by oneself, show a clear positive $T_{C}$ trend. Our analysis also reveals that the B component is richer in metals than its companion ($\sim$0.051 dex), which would indicate that planet host component A has clear chemical differences relative to its companion, which are not expected in a pair of co-eval and co-natal stars. Distinctly, and contrary to some previous results, this is the first $T_{C}$ trend detected in a wide binary system hosting a massive super-Jupiter planet.

We explored some possible scenarios that could originate the observed $T_{C}$ trend in the pair HD\,196067-68. While other alternative causes such as dust cleansing or GCE effects are essentially cancelled out thanks to the co-natal and co-eval nature of this twin-binary system. A possible scenario requires the engulfment of refractory-rich material by the star B. In this case, the possible migration of a hypothetical giant planet could result in the ingestion of refractory material from the inner regions or from the core of the giant planet. This is supported by the general agreement with the predictions from \textsc{terra} (Figure \ref{Figenine}). 
However, the difference in Li {\sc i} observed in HD\,196067-68 could be also explained by considering the difference in stellar ages, temperatures, masses (Figure \ref{Figten}) and metallicities. On the other hand, the sequestrated refractory elements from the A component could have been locked up in rocky planets and/or inside of the core of its planet, as suggested by M09. However, at the moment, only a Jupiter-like planet has been detected around the A component \citep[][]{2013A&A...551A..90M}, while the presence of a core rich in refractory elements on this planet can be possible. In addition, following the results from evolutionary models of \citet[][]{2020MNRAS.493.5079B}, the dust-gas separation in the proto-planetary disk as a consequence of the early formation of a distant giant planet, could also be the origin of the deficiency in refractory elements shown by the A component.

Finally, our analysis suggests that planet accretion, planetary locking, and gas-dust segregation are all plausible. Currently, none of these scenarios can be definitively accepted or rejected if some of the assumptions outlined above are considered. Future high-precision photometric and/or spectroscopic follow-up of the stellar components could provide relevant information over the presence of possible additional planets in this binary system. We also encourage further high-precision chemical studies on new planet-hosting binary systems, which are needed to improve the current knowledge of the star-planet connection. In particular, it would be important to continue studying the possible role of the planetary mass in the chemical differences and mutual $T_{C}$ trends.

\section*{Acknowledgements}
PM and JA acknowledge the financial support from CONICET in the form of doctoral fellowships. J.Y.G. acknowledges the support from a Carnegie Fellowship. MJA thanks the financial support of DIDULS/ULS through the PAAI2021 project. EJ acknowledges funding from CONICET via the PIBAA ID-73669 project. The authors also thank the valuable contribution of Drs. R. Kurucz and C. Sneden for making their codes available to the community. Finally, we warmly thank the anonymous referee for her/his careful reading and valuable comments, 
which allowed us to greatly improve the quality of the manuscript.

\section*{Data Availability}
The spectra collected and analysed in this paper are publicly available at the ESO Archive ( \url{http://archive.eso.org/eso/eso_archive_main.html}).



\bibliographystyle{mnras}

\bibliography{thebibliography}







\bsp	
\label{lastpage}

\end{document}